\documentclass[12pt]{article}
\usepackage{epsfig}
\usepackage{amsmath}

\def\beq{\begin{equation}}
\def\eeq#1{\label{#1}\end{equation}}
\def\eeqn{\end{equation}}

\def\beqa{\begin{eqnarray}}
\def\eeqa#1{\label{#1}\end{eqnarray}}
\def\eeqan{\end{eqnarray}}

\let\bar=\overbar

\def\Dslash{\not{\hbox{\kern-4pt $D$}}}
\def\dslash{\not{\hbox{\kern-2pt $\del$}}}

\def\BR{\mbox{\rm BR}}

\def\msb{{\bar{\ssstyle M \kern -1pt S}}}

\input babarsym
\def\Btag	{\ensuremath{B_{\rm tag}}\xspace}  
\def\mes        {\ensuremath{m_{\rm ES}}\xspace}
\def\Eextra     {\ensuremath{E_{\rm extra}}\xspace}
\def\sB         {\ensuremath{s_B}\xspace}
\def\knunu      {\ensuremath{\B\to K^{(*)}\nu\nub}\xspace }
\def\kxnunu     {\ensuremath{\B\to K\nu\nub}\xspace }
\def\kpnunu     {\ensuremath{\Bp\to\Kp\nu\nub}\xspace }
\def\kznunu     {\ensuremath{\Bz\to\Kz\nu\nub}\xspace }
\def\ksnunu     {\ensuremath{\B\to\Kstar\nu\nub}\xspace }
\def\kspnunu    {\ensuremath{\Bp\to\Kstarp\nu\nub}\xspace }
\def\ksznunu    {\ensuremath{\Bz\to\Kstarz\nu\nub}\xspace }
\def\taunu      {\ensuremath{\Bp\to\taup\nut}\xspace }
\def\lnu      {\ensuremath{\Bp\to\ellp\nul}\xspace }
\def\munu      {\ensuremath{\Bp\to\mup\num}\xspace }

\def\Title#1{\begin{center} {\Large {\bf #1} } \end{center}}

\begin{document}

\Title{{\boldmath $\Bp\!\to\!\taup\nut$ and $B\!\to\! K^{(*)}\nu\nub$ at \babar\ and SuperB}}

\bigskip\bigskip

\begin{raggedright}  
{\it Dana M.\ Lindemann, On behalf of the \babar\ Collaboration}\\
SLAC National Accelerator Laboratory\\
Stanford, California 94309, USA 
\end{raggedright}
\bigskip

\noindent 
{\small {\it Proceedings of CKM 2012, the 7th International Workshop on the CKM Unitarity Triangle, \\
University of Cincinnati, USA, 28 September -- 2 October 2012 }}
\bigskip

\section{Introduction}

The \taunu and \knunu decays are quite distinct from one another, the former a tree-level process and the latter only occurring via loop or box diagrams, and the theoretical motivations for measuring these decays are equally divergent.  However, the experimental search for these two processes is similar, as both decays have final states with a single reconstructed particle and several neutrinos.  Since \taunu and \knunu cannot be fully reconstructed, the clean, hermetic environments of the $B$ Factories provide sensitivity to these decays.  During its lifetime, the \babar\ experiment \cite{ref:babar} collected 429 \invfb of data at the \FourS resonance, which corresponds to $\sim470$ million \BB pairs.  The SuperB experiment \cite{ref:superB}, a next-generation $B$ Factory, aims to produce $75\invab$ at the \FourS resonance over a five-year period.\footnote{Following this conference, the SuperB experiment was canceled. A similar experiment Belle II \cite{ref:BelleII} expects $50\invab$.}

The $\FourS\to\BB$ production at $B$ Factories can be exploited by fully reconstructing one $B$ meson (\Btag) in a number of hadronic final states in order to determine the signal $B$ four-vector for improved resolution on the sighnal kinematics and the missing four-momentum.  The ``energy-substituted" mass of the \Btag, $\mes\equiv\sqrt{E_{\rm beam}^2-{\vec p}_{\Btag}^{~2}}$, peaks at the nominal $B$ mass for correctly reconstructed \Btag candidates, and produces a relatively flat distribution from combinatoric background.  In both the \taunu and \knunu searches described in this paper, the signal efficiency and peaking background, estimated from Monte Carlo (MC) modeling, are validated with data using this \mes distribution.  The combinatoric background is extrapolated directly from data in the \mes sideband regions.

\section{The \taunu measurement}

Within the Standard Model (SM), the leptonic decay \taunu proceeds via an annihilation of $b$ and $u$ quarks into a virtual $W^+$ boson.  Without QCD-based uncertainties in the final state, leptonic decays can provide clean theoretical predictions of SM parameters like $|V_{ub}|$ and the $B$-meson decay constant $f_B$.  Both these parameters dominate the SM uncertainty of the branching fraction
\begin{equation}
\BR(\lnu)_{\rm SM} = \frac{G_F^2 m_B}{8\pi}|V_{ub}|^2f^2_B\tau_B m_{\ell}^2\left(1-\frac{m_{\ell}^2}{m_{b}^2}\right)^2
\label{taunuBF}
\end{equation}
where the square of the lepton mass, $m_{\ell}^2$, indicates a helicity suppression.   Thus, while $\BR(\Bp\to e^+\nue)$ and $\BR(\munu)$ \cite{ref:lnu} have been inaccessible at current $B$ Factories, the SM predicts \BR(\taunu) to be on the order of $10^{-4}$.  This rate can be significantly enhanced or suppressed by the contribution of a charged Higgs boson in place of the $W^+$ boson.  For example, in the Type-II two-Higgs doublet model (2HDM) \cite{ref:2HDM}, Eq.~\eqref{taunuBF} is multiplied by an additional factor $\left(1-\tan^2\beta \frac{m^2_B}{m^2_H}\right)^2$, where $m_H$ is the mass of the $H^+$ and $\tan\beta$ is the ratio of the vacuum expectation values.  Thus, measuring the \taunu branching fraction can constrain the parameters of new physics models.

The recent search for \taunu \cite{ref:taunu}, using the full \babar\ dataset, requires a reconstructed \Btag and exactly one  track corresponding to a one-prong decay of the tau: $e\nu\nub,~\mu\nu\nub,~\pi^+\nu,$ or $\rho\nu\to\pi^+\piz\nu$.  Two event-shape variables are employed to suppress $\epem\to\qqbar~(q=u,d,s,c)$ backgrounds.  In addition, the angle of the missing three-momentum ($\cos\theta_{\rm miss}$) and the momentum of the pion in the center-of-mass frame ($p^{*}_\pi$) are used in a two-variable likelihood ratio to reduce background in the $\tau\to\pi\nu$ channel.  The $\tau\to\rho\nu$ channel uses a four-variable likelihood: $\cos\theta_{\rm miss}$, $p^{*}_{\pi}$, and the invariant masses of the \piz and $\rho$ candidates.  

After reconstructing the \Btag and $\tau$, the sum of the remaining energy in the detector (\Eextra) should be zero.  However, mis-reconstructions and noise typically contribute to additional energy in the event.  Using samples of fully-reconstructed ``double-tagged" events, in which a second $B$ is hadronically or semileptonically reconstructed opposite the \Btag, the MC modeling of \Eextra is validated with data. 

\begin{figure}[htb]
	\begin{center}
	\epsfig{file=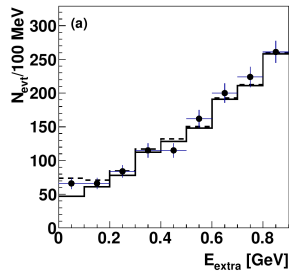,height=1.5in}
	\caption{The final \Eextra distribution in data (points) for all \taunu channels fitted simultaneously.  The best-fit distribution (dashed), which includes the signal excess, is overlaid on the expected background (solid).}
	\label{fig:taunu}
	\end{center}
\end{figure}

The branching fraction is extracted using an unbinned maximum likelihood fit to \Eextra in all $\tau$-decay channels simultaneously.  An excess at low \Eextra is observed, as shown in Fig.~\ref{fig:taunu}, corresponding to an exclusion of the null hypothesis at $3.8\sigma$ (including systematic uncertainties).  This search measures $\BR(\taunu)= (1.83^{+0.53}_{-0.49}\pm 0.24)\times 10^{-4}$ where the uncertainties are statistical and systematic respectively.  Combining this measurement with that of a previous \babar\ semileptonic-tag measurement with an independent dataset \cite{ref:SLtag} gives $\BR(\taunu)= (1.79\pm 0.48)\times 10^{-4}$.

These results are consistent with previous measurements by \babar, all of which suggest tension with other SM measurements.  Assuming $f_B=189\pm4$ \cite{ref:HPQCD}, and extracting $|V_{ub}|$ from either inclusive or exclusive decay measurements, Eq.~\eqref{taunuBF} predicts $\BR(\taunu)_{\rm SM}=(1.18\pm0.16)\times 10^{-4}$ or $(0.62\pm 0.12)\times 10^{-4}$, respectively.  Furthermore, the fit of other experimental values within the Unitarity Triangle expects $\BR(\taunu)_{\rm fit}=(0.832\pm0.84)\times 10^{-4}$ \cite{ref:CKMfit}.  

To determine if this tension is evidence of new physics, the current world precision on $\BR(\taunu)$ of $\sim20\%$ must be reduced.  Simulation studies show that with the $75\invab$ of data expected from SuperB, the $\BR(\taunu)$ precision should improve to about 3--4\% \cite{ref:superB}.  The currently unobserved \munu decay should also be measurable with a 5--6\% precision. In addition to its experimentally-clean two-body final state, \munu would give access to $\BR(\munu)/\BR(\taunu)$, which is independent from the theoretical uncertainties of $f_B|V_{ub}|$.

\section{The search for \knunu}

The flavor-changing neutral current decays \knunu are prohibited in the SM at tree-level, but can proceed via electroweak-penguin or box diagrams.  These rare decays, predicted on the order of $10^{-6}$, are sensitive to new physics scenarios which could contribute at the same order as the SM. These scenarios include non-standard $Z$ couplings, new particles entering into the loops such as from Supersymmetric models, or invisible particles contributing to the final-state missing energy \cite{ref:ABSW}.  In addition to potentially significant branching-fraction enhancements, some new physics models suggest that the \knunu decay kinematics would also show modifications.

A recent (2012) search for \knunu uses the full \babar\ dataset to select events with a reconstructed hadronic \Btag candidate, one additional $K^{(*)}$, and no extra tracks.  This kaon is reconstructed in six signal channels: $\Kp$, $\KS\to\pip\pim$, $\Kstarp\to\Kp\piz$, $\Kstarp\to\KS\pip$, $\Kstarz\to\Kp\pim$, and $\Kstarz\to\KS\piz$.  Six event-shape variables are employed in a likelihood ratio to further suppress $\epem\to\qqbar$ backgrounds, and \Eextra is required to be less than a few hundred MeV.    

\begin{figure}[htb]
	\begin{center}
	\epsfig{file=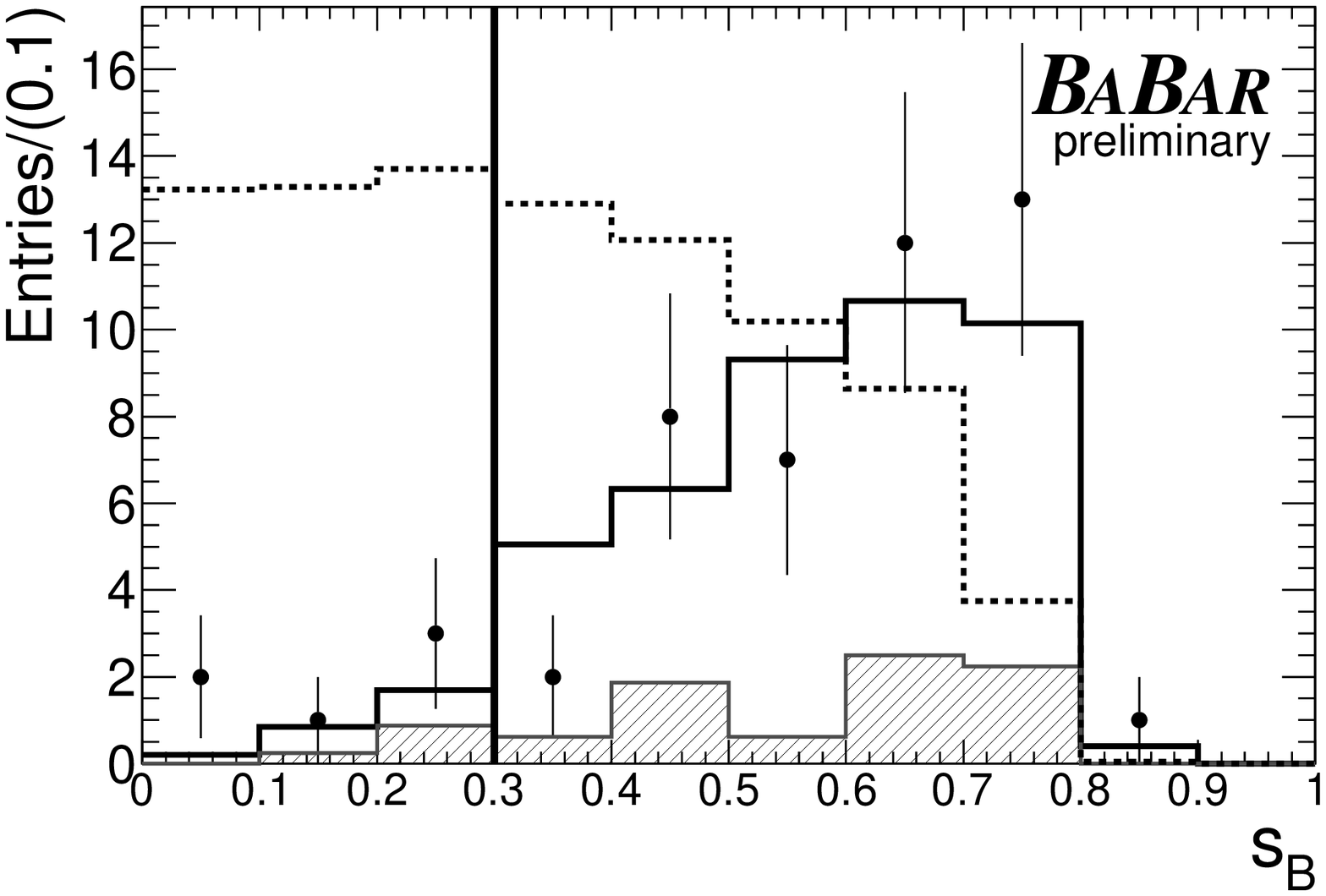,height=1.5in}	
	\hfill
	\epsfig{file=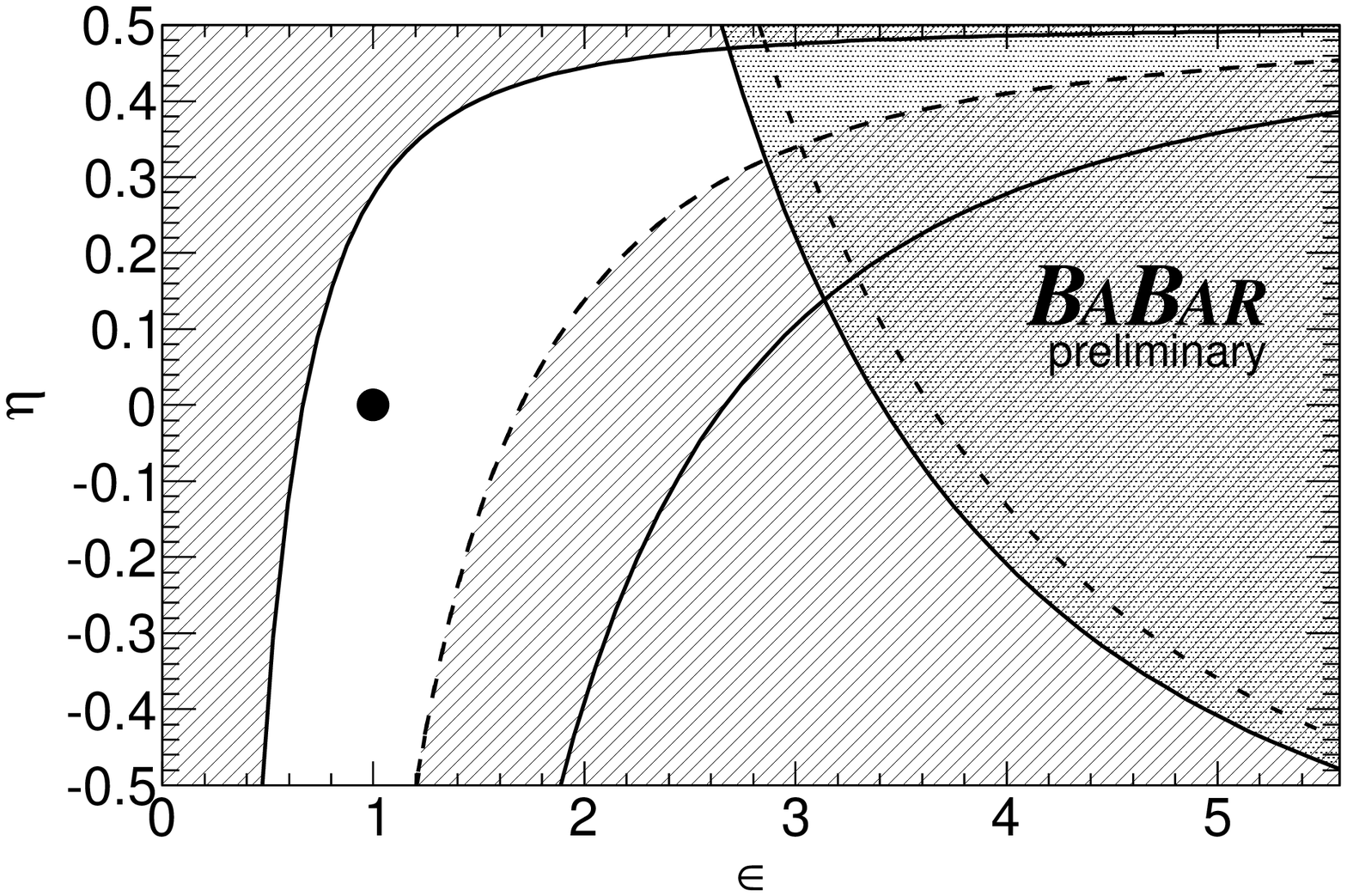,height=1.5in}
	\caption{(a) The final \sB distribution in data (points) for \kpnunu events. The combinatoric (striped) plus \mes-peaking (solid) background are overlaid with signal MC (dashed) normalized to $20\times10^{-5}$ for visibility. The vertical line depicts the signal region. (b) The new physics constraints at 90\% CL on $\eta$ and $\epsilon$.  The \kxnunu (striped) and \ksnunu (shaded) excluded areas are determined from this \knunu analysis (solid lines) and from previous semileptonic-tag analyses (dashed lines).  The dot shows the expected SM value.}
	\label{fig:knunu}
	\end{center}
\end{figure}

The branching-fraction upper limits are obtained within the low $\sB\equiv q^2/(m_Bc)^2$ region, where $q^2$ is the invariant mass of the neutrino pair. This signal region, which corresponds to that of high-momentum kaons, has relatively little background, as shown in Fig.~\ref{fig:knunu}a.  At 90\% confidence level (CL), preliminary upper limits are determined to be $\BR(\kpnunu)<3.7\times10^{-5}$, $\BR(\kznunu)<8.1\times10^{-5}$, $\BR(\kspnunu)<11.6\times10^{-5}$, $\BR(\ksznunu)<9.3\times10^{-5}$, withcd ba	 combined limits of $\BR(\kxnunu)<3.2\times10^{-5}$, and $\BR(\ksnunu)<7.9\times10^{-5}$.  In addition, preliminary lower limits at 90\% CL are set at $\BR(\kpnunu)>0.4\times10^{-5}$ and $\BR(\kxnunu)>0.2\times10^{-5}$, although the significance above the null hypothesis is less than two.  This search provides the most stringent upper limits for \kznunu and \ksnunu with the hadronic-tag reconstruction method.

Since some new physics models suggest enhancements in the kinematic spectrum at high \sB, model-independent sensitivity to new physics is achieved by removing the \sB requirement and determining partial branching fractions over the full phasespace.  This search sets branching-fraction upper limits at 90\% CL for several new physics models at an order of $10^{-5}$.  In addition, several new physics models could affect the Wilson coefficients, $C^\nu_L$ and $C^\nu_R$, where the latter is zero in the SM.  By redefining $C^\nu_{L,R}$ as \cite{ref:ABSW}
\begin{equation}
\label{KnunuEps}
\epsilon \equiv \frac{\sqrt{|C_L^{\nu}|^2+|C_R^{\nu}|^2}}{|C_{L,{\rm SM}}^{\nu}|}~,~~~
\eta \equiv \frac{-{\rm Re}(C_L^{\nu}C_R^{\nu*})}{|C_{L}^{\nu}|^2+|C_{R}^{\nu}|^2}~,
\end{equation}
one can combine the results of \kxnunu and \ksnunu to constrain new physics, as shown in Fig.~\ref{fig:knunu}b.  The \kxnunu lower limits provide the first upper limits on $\eta$ and first lower limits on $\epsilon$.  The excluded parameter-space is in agreement with the SM expectation of $\eta = 0$, $\epsilon = 1$.

The rare decays \knunu may be inaccessible at current $B$ Factories, but they are expected to be measured at SuperB.  Simulation studies show that \kxnunu may be observed at $3\sigma$ with about 10\invab, and \ksnunu around 50\invab \cite{ref:superBstudies}.  With 75\invab and assuming SM values, the precision on the branching fraction measurements are expected to be 15--20\%.  In addition, the longitudinal polarization fraction of \ksnunu decays may be accessible around 75\invab, giving yet another handle on new physics constraints.

\section{Conclusions}
\babar\ has recently measured $\BR(\taunu)$ and obtained limits on $\BR(\knunu)$ using hadronic-tag reconstruction of the recoiling $B$ meson.  The result $\BR(\taunu)= (1.83^{+0.53}_{-0.49}\pm 0.24)\times 10^{-4}$ is consistent with previous \babar\ measurements, although it is slightly above the SM expected value.  The rare decays \knunu are not yet observed, but the new upper and lower limits are approaching the SM predictions. With the expected data from SuperB, the \taunu tensions will likely be confirmed or excluded with improved precision, and \knunu and \munu decays are expected to be observed.

\end{document}